\documentclass[prb,showpacs,showkeys]{revtex4}% Physical Review B

\usepackage{graphicx}% Include figure files
\usepackage{dcolumn}% Align table columns on decimal point
\usepackage{bm}% bold math

\begin{document}

\preprint{APS/123-QED}

\title{The conceptualization of time and the constancy of the speed of light}% Force line breaks with \\

\author{Vasco Guerra}
 \email{vguerra@alfa.ist.utl.pt}
% \affiliation{Departamento de F\'{\i}sica}
 \altaffiliation[Also at ]{Centro de F\'{\i}sica dos Plasmas, Instituto Superior T\'{ecnico}, 1049-001 Lisboa, Portugal}
\author{Rodrigo de Abreu}%
\affiliation{%
Departamento de F\'{\i}sica, Instituto Superior T\'{ecnico}, 1049-001 Lisboa, Portugal
}%

\date{\today}% It is always \today, today,

\begin{abstract}

In this work we show that the null result of the Michelson-Morley experiment in vacuum is deeply connected 
with the notion of time. It can be deduced without any mathematics only from the assumption that all good clocks 
can be used to measure time with the same results, independently of the machinery involved in their manufacturing.

A second important assumption, intrinsic to the very notion of time, is that clocks
measure time in the same way in different frames, \textit{i.e.}, the notion of time is the same in all inertial frames.
Under this assumption, we point out that the ``postulate" of constancy of the ``two-way" speed of light in vacuum in 
all frames independently of the state of motion of the emitting body is also strongly related to the concept of time,
together with the existence of a limit speed in the ``rest frame". This postulate simply results from the construction of clocks
where tic-tacs are made by objects traveling with the limit speed.

\end{abstract}

%Use showkeys class option if keyword
\maketitle

\section{Introduction}

Einstein's celebrated article ``On the electrodynamics of moving bodies" \cite{Einstein1905} was published precisely 100 
years ago. During this time, Special Theory of Relativity established itself as one of the most exciting topics in Physics. 
The challenges and results of Special Relativity are so stimulating that it keeps attracting the attention of physicists 
and philosophers, fascinating the general public as well. The centenary of Special Relativity and the 2005 World Year of 
Physics provide the perfect occasion to revisit its foundations. In this work we suggest that the roots of both
the null result of Michelson-Morley experiment and of the postulate of the constancy of the speed of light in vacuum
in all inertial frames, independently of the state of motion of the 
emitting body, rest firmly on the very notion of time.

The structure of this article is as follows. In the next section we very briefly describe Michelson-Morley
experiment, to somehow introduce the scenario. Section \ref{sec3} deals with the notions of ``time" and ``clocks".
Finally, section \ref{sec4} contains the main discussion, proposing that the null result of the Michelson-Morley experiment and 
the postulate of the constancy of the speed of light in vacuum are strongly related with the concept of time.

\section{The Michelson-Morley experiment}

\begin{figure}
\begin{quote}
\begin{center}
\includegraphics[width=8cm]{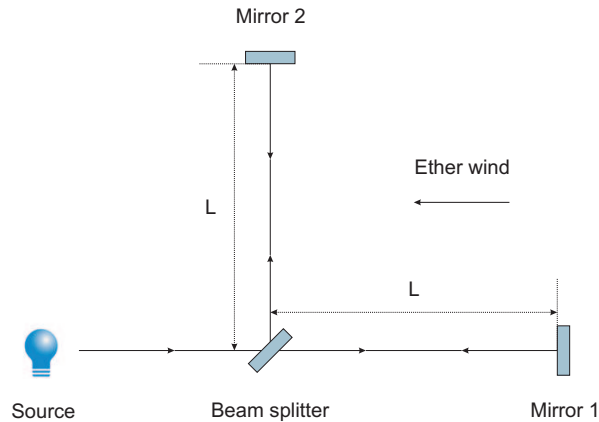}
\end{center}
\caption{Schematic diagram of the Michelson-Morley experiment.}\label{fig1}
\end{quote}
\vskip -0.5cm
\end{figure}

As it is well known, in the end of 19$^{th}$ century
several scientists admitted that light waves move through a light ether and that the speed of light in vacuum
was $c$ only in a special, absolute frame at rest with respect to this ether. 
Now, if light moves with speed $c$ only with respect to one special frame,
it was supposed the speed of light on earth should be faster or slower than $c$,
depending on the way the earth would be moving through the ether.
Several attempts were made to determine the absolute velocity
of the earth through the ether. The most famous was performed by Michelson and Morley
in 1887.
A simplified scheme of the Michelson interferometer is shown in figure \ref{fig1}.
Essentially a light source emits a beam of light which is divided at a
beam splitter. The two resulting beams continue in perpendicular directions
to mirrors 1 and 2, where they are reflected, coming back to the
same point, where they are recombined as two superposed 
beams. The details of the experiment are not too complicated to follow and can be
found in any physics textbook, such as the classic ones by Feynman \cite{Feyetal1979}
or Serway \cite{SB2000}.
Basically, what happens is that
if the time taken for the light to go from the beam splitter to mirror 1 and back 
is the same as the time from the beam splitter to mirror 2 and back, then the two
beams would reinforce each other. However, if these times differ slightly, an interference
pattern should be formed. If the interfermoter is at rest (\textit{i.e.}, on the ether frame) and in vacuum, the times should be 
precisely equal. But if it is 
moving with a certain speed it was expected they would be different. Yet no significant time difference was found: 
it seemed the speed of the earth through the ether could not be detected.
The small time differences found were considered at the time to be merely errors of experiment and
it was concluded the experiment had given a null result.
This was of course a puzzling issue, that was solved by
Lorentz in 1895 \cite{Lorentz1895}. 
He suggested that all moving bodies contract in the direction paralel to their
movement through the ether. Lorentz has shown that if the length of a moving
body is contracted by a factor $\sqrt{1-v^2/c^2}$, 
and this contraction occurs only in the direction of the motion, then
the null results of Michelson-Morley experiment would be readily explained.
A rather similar hypothesis of a change in the 
length of material bodies had been formulated independently by Fitzgerald in 
1889 \cite{Fitzgerald1889}, who promoted his deformation idea in his lectures and correspondence.
Einstein's Special Relativity ``solves" the problem of the null result
of Michelson-Morley experiment by postulating that the speed of light is constant and independent of the
speed of the source not only in a preferred ether frame, but in \textit{all} inertial frames.
The Michelson-Morley experiment is thus related to the postulate of the constancy of the speed of light 
in vacuum and plays a central 
role in Special Relativity, appearing in almost all books presenting the subject.

\section{Time and clocks}\label{sec3}

We all have an intuitive idea about the notion of time.
However, it is likely we will have an hard time in answering in a simple way the
fundamental question ``what is time?".
Many scientists and philosophers have reflected about this question. One celebrated
example is that of St. Augustine, who wrote in his ``Confessions":
\begin{quote}
What is time? Who can readily and briefly explain this? Who can even in thought comprehend it, 
so as to utter a word about it? But what in discourse do we mention more familiarly and 
knowingly, than time? And, we understand, when we speak of it; we understand also, when we 
hear it spoken of by another. What then is time? If no one asks me, 
I know: if I wish to explain it to one that asked, I know not.
%: yet I say boldly that I know, that if 
%nothing passed away, time past were not; and if nothing were coming, a 
%time to come were not; and if nothing were, time present were not. 
\end{quote}
Interestingly enough, the difficulty in defining in words what is time is not of 
great concern to start studying physics... since we know how to measure it! 
Time is what is measured with clocks. Therefore, the ``only" things needed to proceed are clocks. 

Let us for now follow Einstein \cite{Einstein1905} and restrict the analysis to 
\begin{quote}
a coordinate system 
in which Newton's mechanical equations are valid.
To distinguish this system verbally from those to be introduced later, and to make our
presentation more precise, we will call it the ``rest system."
\end{quote}
We will further confine the discussion to clocks in ``vacuum", in order to avoid the problem
of the interactions between clocks and the surrounding medium.
In principle, 
any periodic phenomenum may be associated with a clock. Galileo even used the rhythm of his
heart beat as a clock. Which raises the question of how do we know if the time intervals
given by a certain clock are really equal. The truth is that we do \textit{not} know.
What is possible to do is to compare the readings of different clocks (see below). Then, using these
comparisons and with the help of theoretical arguments about the laws ruling each of the
periodic phenomena involved, decide which clock is more trustful. 

\begin{figure}
\begin{quote}
\begin{center}
\includegraphics[width=1.8cm]{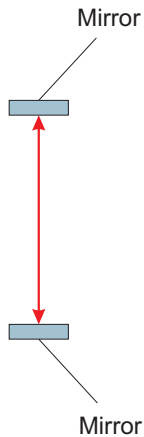}
\end{center}
\caption{The Feynman light clock.}\label{fig2}
\end{quote}
\vskip -0.5cm
\end{figure}

One very simple clock is Feynman's light clock \cite{Feyetal1979}, schematically depicted in 
figure \ref{fig2}.
It consists of two mirrors vertically aligned and a light source close to one of the mirrors. At a certain instant the 
source emits one photon in the direction of the other mirror.
The photon is continuously reflected by both mirrors, making ``tic" each time it is reflected on the upper mirror, and ``tac"
each time it is reflected on the lower one.
The unit of time is then defined by a complete ``tic-tac". Notice that rigourously it is not convenient to define the
unit of time with just the ``tic" or just the ``tac", because this would involve the additional assumption that one is equal
to the other. 
This assumption is generally accepted to hold, but it is by no means obvious nor even necessary, as pointed out 
in \cite{Edwards1963,MS1977}. Such discussion is far beyond the purpose ot the present work and we simply want
to stress that the periodic movement corresponds to the complete ``tic-tac". That being so, for a general definition of time 
this is the time interval that must be considered.

\begin{figure}[h]
\begin{quote}
\begin{center}
\includegraphics[width=8.5cm]{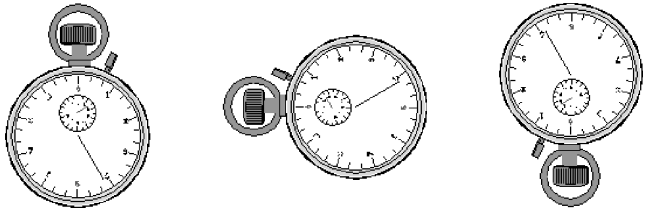}
\end{center}
\caption{Identical clocks with different space orientations measure the same times.}\label{fig3}
\end{quote}
\vskip -0.5cm
\end{figure}

It is implicitly assumed that time passes independently of the type of clocks we use to measure it.
Therefore, we expect that clocks built upon different phenomena, when in the same location, will mark the same times.
This is of course an assumption, 
but even though it is what we expect. In any case, it can be verified by experiment. And up to now there is no reason
to suspect that different clocks in the same place do not provide the same time readings, independently of the machinery involved in 
their construction, as long as they are precise enough.
If this idea is true, if any good clock can be used to measure time,
then the time measurements of a certain clock do not depend on its \textit{orientation}. 
Even if two equal clocks with different orientations may be regarded as two different clocks, they should still
measure the same time intervals (figure \ref{fig3}).
If our notion of time is correct, this must be true not only in the ``rest system", but in any ``moving" inertial frame as 
well, since the clocks used to measure time in these frames are exactly equal to the ones used in the ``rest system".

\section{The constancy of the speed of light}\label{sec4}

\begin{figure}
\begin{quote}
\begin{center}
\includegraphics[width=9cm]{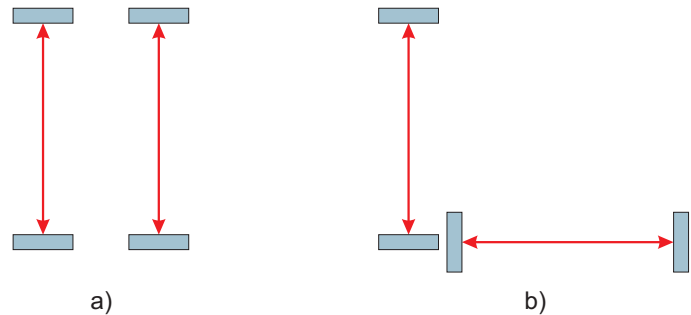}
\end{center}
\caption{Two similar Feynman light clocks give a complete tic-tac exactly at the same instant both when a) they are side by side 
and b) one of them is rotated by 90 degrees.}\label{fig4}
\end{quote}
\vskip -0.5cm
\end{figure}

We are now ready to discuss the Michelson-Morley experiment and the postulate of the constancy of the speed of light.
Consider two Feynman light clocks, exactly equal, placed side by side. The particular periodic movement involved in Feynman clocks 
repeats itself after a complete ``tic-tac", corresponding to a \textit{round trip} of light from one mirror to the other and back. 
The time-unit
is one tic-tac and is the same for both clocks. Now rotate one of the clocks by 90 degrees, as shown in figure \ref{fig4}.
What happens? If time does not really depend on the clocks used to measure it, if time does not depend on the orientation
of clocks, then the complete tic-tac of both clocks is still the same. But two Feynman clocks rotated by 90 degrees
are no more no less than Michelson-Morley interferometer! 
Therefore, the times light takes to go along each arm of the
interferometer and back \textit{must} be the same. 
This must be true for the ``rest system" as well for any ``moving" inertial frame.
In other words, in \textit{each} inertial frame the \textit{two-way} speed of light is constant. 
Once more notice that to keep the argument completely general we must use the two-way speed of light, which corresponds
to the average speed of light when it makes a round trip, since the complete ``tic-tac" defining the time unit involves
such a round trip. It is not necessary to invoke any additional assumption about the \textit{one-way} speeds of light,
which dictate the time intervals for the ``tic"  and the ``tac".

As we have just seen, the null result of Michelson-Morley experiment is actually a confirmation
that any good clock can be used to measure time and is deeply connected with the notion of \textit{time}. 
The null result of Michelson-Morley experiment
and the constancy of the two-way speed of light may seem odd, but it would be a disaster for
our notion of time if it had been otherwise. And, quite surprisingly, the null result of Michelson-Morley
experiment could probably have been anticipated. 

The Michelson-Morley experiment shows that in vacuum light takes the same time to 
go up to the mirrors and back in both arms of the interferometer (see figure \ref{fig1}).
The null result of the experiment is related only to the equality of both these times. It corresponds
to the assertion that in \textit{each} particular inertial frame the two-way speed of light is the
same in all directions. 
This is already some ``constancy" of the speed of light, but the experiment does not tell it has the same 
value in \textit{all} inertial frames.
Lorentz original explanation of the experiment was based only on the space contraction hypothesis
and time was not affected. To arrive at the conclusion that the two-way speed of light is the same
in \textit{all} inertial frames it is necessary to go beyond the null result of Michelson-Morley experiment, by
actually measuring its value. We shall now see that this second kind of ``constancy" is also
implied by the fundamental concept of time.

Imagine a Feynman-like clock with bullets. The argument given above is 
related only with time measurements and equivalence of clocks, not with \textit{light} clocks. 
It can be applied with clocks where bullets travel instead of light rays and guns replace the mirrors.
Therefore, in \textit{each} inertial frame the two-way speed of bullets must be the same, 
otherwise the tic-tacs given by two bullet clocks rotated by 90 degrees would be different.
Is that really so?
Yes it is. But what is different with light is that, contrary to bullets, its two-way speed is the same
in \textit{all} inertial frames, \textit{independently of the speed of the source}. This very striking fact is connected 
to $c$ being a limit speed in the ``rest system", as we will now show.

We have just seen that if time does not depend on the clocks used to measure it, then in \textit{each} inertial
frame the two-way speed of \textit{any} object (emmited by perfectly
equal devices) must be the same. This is true for bullets, as well as for light.
Let us now give a second step, which deals with a different idea of ``constancy" of speed. Consider again a Feynman-like 
clock with bullets. 
We already know that in any particular moving inertial frame $S^\prime$, say a train that is passing in a 
station, the two-way speed of bullets fired in that frame is the same, no matter the direction they are travelling.
Now, if the two-way speed of these \textit{same} bullets, that were shot with 
guns fixed in the train, is measured in the station, one obtains a different value. Thus, the \textit{same} bullets travel 
with \textit{different} two-way speeds in different frames. 
In this sense one can say the two-way speed of bullets is \textit{not} the same in \textit{all} inertial frames.
However, suppose now that the bullet-clock is taken 
from the train to the station and bullets are again shot. It is precisely the same clock, which does ``tic-tac"
and measures time precisely in the same way. 
Notice that \textit{time is actually measured with distances} and that since a clocks involve a periodic movement,
time units -- the tic-tacs -- are defined with average two-ways speeds.
Thus, if bullets move in $S^\prime$
with a certain two-way speed $v^\prime$ along a straigth line, the total distance $x^\prime$ they travel in 
a round trip in $S^\prime$
is related with the elapsed time $t^\prime$ via $x^\prime=v^\prime t^\prime$. For instance, if the round trip
takes 1 meter and the two-way speed is 1 meter per second, one second passes in each tic-tac.
If the same clock is used in a second inertial frame $S^{\prime\prime}$, in each tic-tac bullets now travel
a distance $x^{\prime\prime}$, which corresponds to a certain time $t^{\prime\prime}$.
Of course $x^{\prime\prime}=v^{\prime\prime}t^{\prime\prime}$, where $v^{\prime\prime}$ is the
two-way speed of bullets in $S^{\prime\prime}$. But, if the clock measures time in the same way as before, 
then this correspondence between distance and time must be
done exactly in the same way, \textit{i.e.}, $v^\prime=v^{\prime\prime}$ and the two-way speeds of bullets 
must be the same in both frames.
Hence, the two-way speed, measured in the station, of bullets fired in the station, must be the same as the two-way speed, 
measured in the train, of bullets fired in the train. This is intrinsic to the very notion of time: it only means clocks
measure time in the same way in different frames: the notion of time is the same in all inertial frames.
So there is also a ``constancy" of the 
two-way speed of bullets in \textit{all} inertial frames! But this constancy refers to \textit{different} objects, the bullets fired 
in the train and the bullets fired in the station.
\textit{Different} bullets travel with the \textit{same} two-way speed in different frames. 
Evidently the argument is valid not only for bullets. Therefore, the two-way speeds of \textit{any} kind of 
objects must be the same in \textit{all} inertial frames, as long as we are referring to \textit{different}
objects of the same kind, emmitted by perfectly equal devices at rest in relation to the different frames.
This is true for bullets, as well as for light.

The final step is the issue of a limit speed. Assume that such a limit exists: that no object travelling in the ``rest system" 
can have a speed higher than a certain value $c$.
To all objects moving with speeds lower than the limit speed,
the constancy of their two-way speed in all frames refers to \textit{different} objects, such as the ``bullets fired in the train"
and the ``bullets fired in the station". For simplicity, let us make the station coincide with the ``rest frame" and refer
speeds to this frame. If the bullets
are fired in the direction of the head of the train, bullets shot from the train always have a higher speed than
the bullets shot from the station with a perfectly equal gun.
Now pick a more powerful gun, that shoots bullets at higher speeds. The same thing happens. The bullets shot from the train
always have a higher speed than the bullets shot from the station, which, in turn, move faster than the bullets shot
with the less powerful gun. The process can continue by successively taking more powerful guns. For a really powerful gun,
the speed of bullets fired in the station can be very close to the limit speed. But in the ``rest frame" of the station
the bullets
fired from the moving train cannot have a speed higher than the limit speed! Therefore, in this case the bullets 
shot from the train will move almost imperceptibly faster than the bullets shot from the station. And so, if an object
emmited from the ``rest frame" moves with the limit speed $c$, it will also move with speed $c$ when emmitted from a
``moving" inertial frame: \textit{objects travelling with the limit speed must do so independently of the velocity of the 
emmiting source}.
In this way, the existence of ``objects" moving with the same speed independently from the speed of 
the source is directly connected to the existence of a limit speed in the ``rest frame".

What is special about light is that it moves with the limit speed. That being so, and contrary to bullets, if two light rays are 
emitted simultaneously, on the same position in space, one in the train and the other in the station, there is no difference between 
the two rays! Light propagates on the ``rest frame" with the same speed independently of the velocity of the source emitting 
the light, and actually both rays can be seen as forming the \textit{same} object. In the ``second step" it was argued
that different objects of the same kind have the same two-way speed in different frames. This is true for bullets, as well as for light.
But with light the distinction between ``different" light rays is artificial. The two-way speed of the \textit{same} light rays 
must then be constant in different inertial frames. And this speed must be the limit speed in the ``rest frame". 
Because different objects, emmited from different inertial 
frames, collapse into the same object if and only if they travel with the limit speed $c$. Only in this case the
constancy of their two-way speeds refers to the same object.
The \textit{same} light rays travel with the \textit{same} two-way speeds in all 
inertial frames. 
This distinctive feature of the limit speed, that its two-way value is the same
in all inertial frames regardless of the speed of the source, makes it the privileged way to convert spaces to times through
$x=ct$. All the argumentation was made only on the basis of our notions of time and clocks. In this sense, the limit speed can 
somehow be seen as ``the speed of time".

\section{Summary and conclusion}

We have shown that both the null result of Michelson-Morley experiment and the postulate of constancy of the two-way
speed of light in vacuum are a direct consequence of the fundamental notions of time and clocks. They can be obtained under 
three very reasonable assumptions: i) that all good clocks can be used to measure time, independently of the periodic physical phenomena
they are built upon; ii) that time is measured in the same way in all inertial frames, \textit{i.e.}, 
if a particular clock can be used to measure time in the ``rest system", a similar clock can be used to measure time
in ``moving" inertial frames; iii) that a limit speed exists in the ``rest system". 

The question of the possible constancy of 
the one-way speed of light in vacuum is related to the one century long
question of a preferred frame \textit{vs.} ``equivalence" of all inertial frames, and is left for a forthcoming paper.
However, the arguments presented here are completely general and have be true in both scenarios.

%\begin{acknowledgments}
%We wish to acknowledge the support of the author community in using
%REV\TeX{}, offering suggestions and encouragement, testing new versions,
%\end{acknowledgments}

%\newpage %Just because of unusual number of tables stacked at end
\bibliography{vasco}% Produces the bibliography via BibTeX.

\end{document}